# Enhanced electrocatalytic oxygen evolution activity in geometrically designed SrRuO$_3$ thin films

Abhijit Biswas[a,†], G. Shiva Shanker[a,†], Tisita Das[b], Rajesh Mandal[a], Sudip Chakraborty[c,*], and Satishchandra Ogale[a,d*]

[a] *Department of Physics and Centre for Energy Science, Indian Institute of Science Education and Research (IISER) Pune, Pune, Maharashtra, India-411008*
[b] *Department of Materials Science, Indian Association for the Cultivation of Science, Jadavpur, Kolkata, India-700032*
[c] *Discipline of Physics, Indian Institute of Technology (IIT) Indore, Simrol, Indore, India-453552*
[d] *Research Institute for Sustainable Energy (RISE), TCG Centres for Research and Education in Science and Technology (TCG-CREST), Kolkata, India-700091.*

**ABSTRACT**

For generation of sustainable, clean and highly efficient energy, the electrocatalytic oxygen evolution reaction represents an attractive platform, thus inviting immense research activities in recent years. However, designing the catalyst with enhanced electrocatalytic activity remains one of the major challenges. Here, we examined the oxygen evolution reaction activities of geometrically designed (with and without step-textured morphology) thin films of an electrocatalytically active correlated metallic SrRuO$_3$ perovskite grown on *c*- and *r*-plane sapphire substrates. On *c*-plane sapphire, as compared to the uniform surface, the step-textured films endowed with active Ru-sites show remarkable decrease in the overpotential (~25 mV). Interestingly, the behavior is opposite for the *r*-plane case, highlighting the significance of the active sites, in addition with the polar surface termination of selective crystal facets. Density functional theory calculation confirms the favorable energy reaction pathway for the active site dependent enhancement in OER. Our strategy might pave the way towards designing the surfaces of various oxide thin films for high performance energy conversion based devices.

[†] These authors contributed equally to this work
[*] Corresponding authors: satishogale@iiserpune.ac.in, satish.ogale@tcgcrest.org, sudip@iiti.ac.in

**1. Introduction**

The demand for enhanced use of green and sustainable energy is increasing continuously due to the significant negative impact of polluting fuels on the environment. One of the most sought after clean fuel is hydrogen and its efficient and economical production is a major challenge [1,2]. One of the key components of this process is the oxygen evolution reaction, and in the alkaline electrolysis process (which is favored over the unstable acid electrolysis) the oxygen evolution reaction (OER) kinetics is very sluggish because it is a four electron-proton coupled reaction ($4OH^- = O_2 + 2H_2O + 4e^-$) that requires high energy to overcome the kinetic barriers. Accordingly, the overpotential are quite high inviting research efforts to reduce their values by novel catalytic design. Fundamentally, the OER activity depends upon various parameters e.g. crystal structure, surface termination, surface energy, surface oxygen binding energy, valence state(s), surface morphology, electrical conductivity, and importantly the presence of active sites [1,2].

Therefore, in recent years extensive efforts have been expended to understand the origin of OER activity in the special class of environment friendly transition metal based $ABO_3$ perovskite catalysts [3]. These are ruthenates (Ru-based)[4,5], iridates (Ir-based)[6] nickelates (Ni-based)[7], manganites (Mn-based)[8] and cobaltates (Co-based)[9] wherein most physical properties are determined by the $BO_6$ octahedral architecture, whereas the A-site cation controls the conduction band energy and the degree of deviation from the ideal crystal structure. These oxides show promising energy conversion activities along with its ample emergent physical phenomena. However, the origin and enhancement of their electrochemical energy conversion performance remains challenging due to the complex interplay between various parameters that hinders their practical implications. Matsumoto *et al.*, proposed that OER takes place on the surface of perovskite oxides, when the B-site transition metal forms the σ* band within the lattice, thus attaining a high oxidation state which would lead to the enhanced catalytic activity [10]. Suntivich *et al*. showed that intrinsic OER activity exhibits a volcano-shape dependence on the occupancy of the *d*-electron with an $e_g$ symmetry of surface transition metal cations [11]. Another report by Bockris and Otagawa showed that electronic structure and the corresponding bond-strength of the surface oxygenated intermediates as a descriptor is important for the OER activity [12]. Man *et al.*, did exhaustive theoretical calculations proposing a fundamental relationship between the catalytic activity and the binding energies of the surface intermediates (O* and OH*) and suggested a wide-range of perovskite catalysts for the OER [13]. The volcano-type plot showed by Reier *et al*., clearly suggests that exposure of B-site cations (e.g. Fe, Co, Ni, Ru and Ir) with high valence state in the electrocatalytically active transition metal based perovskite oxides should show excellent OER activity [14]. However, in most of these studies including those of other classes of catalysts primarily uniform (flat) surfaces are used. Since non-ideal (e.g. step-textured) surfaces provide a plethora of morpho-chemically rich array of catalytic (kink) sites, it is extremely interesting to harness such a possibility through a simple experimental design that can elucidate the importance of the attendant parameters. This is indeed the objective of the present study.

In this report, by geometrically designing a well-known correlated metallic $SrRuO_3$ thin film on *c*- and *r*-plane sapphire substrates, we have addressed the significance of the active sites in conjugation with crystal facet, with different atomic configurations. We have found remarkable enhancement in OER with the significant reduction of the overpotential (~25 mV) only in the case of *c*-plane sapphire engineered with step-textured geometry, as compared to uniform (flat) film. In contrast, interestingly an opposite trend has been found for film on *r*-plane sapphire, which implies that in addition to the possibility of enhanced binding enabled by higher chemical coordination afforded by active sites at the steps, specific morph-chemical nature of the step edges governs the phenomenon. Density-functional theory calculations are also performed to elucidate the specific role of active sites for the enhancement.

Structurally, at ambient conditions $SrRuO_3$ forms a $GdFeO_3$ type distorted orthorhombic structure with lattice constants $a$~5.5670 Å, $b$~5.5304 Å, and $c$~7.8446 Å, thus pseudo-cubic with lattice constant ($a_{pc}$) ~3.93 Å. It is a correlated ferromagnetic metal with room temperature resistivity ~200 μΩ·cm and $T_c$ ~160 K [15]. Its strong electron correlation and metallic conductivity (originating from Ru) makes $SrRuO_3$ one of the strong OER-active material [4, 5, 16]. Recently, several reports have shown that increased catalytic activity of $SrRuO_3$ depends on the crystal facets, doping as well as on the modification of the crystal structure [4, 5, 16-19]. Efforts have also been made to stabilize the $SrRuO_3$ catalyst by using a capping layer, as it dissolves rapidly (in both acidic and basic media) at the onset of OER due to the transformation of $Ru^{4+}$ into $Ru^{>4+}$ [4].

## 2. Material and methods
### 2.1. Experimental details

We first grew ~40 nm $SrRuO_3$ thin films on both *c*- and *r*-plane sapphire ($Al_2O_3$, hexagonal structure) substrates by pulsed laser deposition and measured their OER activities. Films were grown by pulsed laser depositions (PLD) method on hexagonal *c*- and *r*-Al$_2$O$_3$ substrates by using the following growth conditions: growth temperature ~700 °C, laser fluence ~3 J/cm$^2$, repetition rate of 2 Hz and oxygen pressure (P$_{o2}$) of 100 mTorr. After the deposition, films were cooled down at the same oxygen pressure to reduce the oxygen vacancies. For achieving the desired step-edge surface geometry, we used two differently shaped masks (flat spade shape; size 4×4 mm$^2$ and square shape, size 1.5×1.5 mm$^2$), made of high purity stainless steel. Before using, we cleaned these masks ultrasonically in high purity acetone (99.9%) for 30 min. After growing the designed uniform (spade-shaped) film for bottom contact, we kept the square box-shaped mask on the top of it to create the vertical wall of step-edge active sites, defining the measurement area.

X-ray diffractions were performed with a Bruker D2 PHASER X-ray Diffractometer. Atomic force microscopy (AFM) was used for the surface topography analysis and measuring the thickness of the films, using Nanosurf AFM (Switzerland). Electrical measurements were performed in standard AC transport four-probe method by using a Quantum Design Physical Property Measurement System (PPMS), within the temperature range of $T$ = 5-300 K. Field Emission Scanning Electron Microscope (FESEM) imaging was done using Zeiss Ultra Plus 4095 operating at 20 kV. X-ray photoelectron spectroscopy (XPS) was done with a K$_\alpha$ X-ray Photoelectron Spectrometer (Thermo-fisher Scientific Instrument, UK). It was carried out in an ultra-high vacuum chamber (2×10$^{-9}$ mBar) by using an Al K$_\alpha$ x-ray source with 6 mA beam current (beam spot size on the sample was ~400 μm).

All the electrochemical measurements were carried out using CH Instruments Potentiostats with a conventional three-electrode system using platinum wire as a counter electrode, Hg/HgO as the reference electrode, and SrRuO$_3$ thin films (catalysts) as a working electrode, respectively. The electrochemical experiments were performed in 1 M KOH electrolyte solution after the purging N$_2$ gas without iR correction. Afterward, the potentials were converted into reversible hydrogen electrode (RHE) by following equation.
$E_{RHE}$ = E vs Hg/HgO + 0.059 × pH + $E^0$ (Hg/HgO), where $E^0$ (Hg/HgO) = 0.098 V.

## 2.2 Theoretical simulations

All the geometry optimization and electronic structure calculations are carried out using Density-functional theory (DFT) based Vienna ab initio simulation package (VASP) [20,21]. The model supercells are periodic along X and Z directions. A vacuum of ~15 Å normal to the surface has been used in order to avoid the interaction between two adjacent images. Throughout the electronic structure calculations, the exchange-correlation functional has been approximated using Perdew-Burke-Ernzerhof (PBE) type generalized gradient approximation (GGA) [22]. The Projector Augmented Wave (PAW) method has been used to describe the ion-electron interaction with a converged kinetic energy cut-off of 500 eV [23]. In order to find the minimum energy configuration, all the structures are fully relaxed until the Hellman−Feynman forces of the constituent atoms become smaller than 0.01 eV/Å, while self-consistency has been achieved with a 0.001 eV/Å convergence accuracy. The Brillouin zone sampling has been done using 5×1×5 Monkhorst-Pack k-points grid during the geometry optimization and total energy calculation [24]. For all the surface calculations, we have considered DFT-D2 type Dispersion corrections as proposed by Grimme [25]. The model structure is constructed by expanding $SrRuO_3$ (111) unit cell to a 2×1×2 supercell in order to be consistent with the experimental findings. The bottom two layers are frozen and the remaining top three layers are allowed to relax, where one Sr-O atom sheet is sandwiched between two Ru-O atom sheets. In order to construct the OER pathway we have calculated the adsorption energy of the reaction intermediates OH*, O* and OOH* on top of both the uniform as well as step-textured $SrRuO_3$ surfaces. The reaction intermediates are most likely to adsorb on top of surface Ru atom.

## 3. Results and discussion

From crystallographic and lattice symmetry standpoint, in case of a cubic (pseudo-cubic) crystal, the film growth on hexagonal $c$-$Al_2O_3$ substrate should favor (111) orientation growth. In principle, two neighboring (111) planes of a cubic $ABO_3$ oxide forms "buckled' honeycomb lattice with trigonal lattice geometry. The x-ray diffraction (XRD) data for the $SrRuO_3$ films reflect (111) growth direction (on $c$-$Al_2O_3$) and along (110) direction (on $r$-$Al_2O_3$) (Fig. 1(a), (b)). AFM shows uniform film with roughness of ∼2.80 nm (on $c$-$Al_2O_3$) and 5.62 nm (on $r$-$Al_2O_3$) (Fig. 1(c)). Stoichiometry of both $c$- and $r$-cut films were obtained by Rutherford backscattering spectrometry (RBS) showing ∼1:1 ratio of Sr and Ru. Interestingly, the sheet resistance ($R_S$) of the film on $c$-$Al_2O_3$ substrate is lower than the film on the $r$-$Al_2O_3$ substrate (Fig. 1(d)), highlighting the vital role of structurally anisotropic transport in $SrRuO_3$ films [26].

Atomically, $SrRuO_3$ along (111) shows alternate stacking of polar termination layers of $Ru^{4+}$ and $SrO_3^{4-}$ (Fig. 2(a)). Consequently, it has high surface energy due to non-polar surface terminations, a crucial factor for the enhancement in the OER activity. In contrast, (110) orientation of SrRuO$_3$ shows alternating polar termination layers of SrRuO$^{4+}$ and O$_2^{4-}$ (Fig. 2(b)). Interestingly, both the (111) and (110) directions have polar stacking, but with alternating layers of Ru$^{4+}$-SrO$_3^{4-}$-Ru$^{4+}$ and SrRuO$^{4+}$-O$_2^{4-}$-SrRuO$^{4+}$, respectively [4]. Clearly, the isolated Ru$^{4+}$ atom termination possibility is finite in the (111) SrRuO$_3$ while the same is absent in the (110) SrRuO$_3$ film. Additionally, the in-plane strain due to lattice mismatch is different for films grown on two different substrates (*c*- and *r*-Al$_2$O$_3$), making both the films of quite different nature, structurally as well as electronically.

Having noted that, we set out to examine the OER activities of these two films. Since it is important to normalize the current density with the film surface roughness, we obtained the roughness values for the two cases from the AFM data. Therefore, the current density was *normalized for both the films with reference to the surface roughness* (henceforth we normalized the current density values duly with their respective film roughness for all the cases) and shown in (Fig. 2(c)). The overpotential (defined as the electrode potential-1.23 V at current density $j$ = 1 mA/cm$^2$) values were found to be ~137 mV (on *c*-cut) and ~217 mV (on *r*-cut); as shown in Fig. 2(d). Now, the roughness for the film on *r*-cut is much higher than that for the film on *c*-cut substrate. However, interestingly, the overpotential value is found to be lower for film on *c*-cut than on *r*-cut substrate. Also, the values of overpotential *without the surface roughness normalization* were found to be ~67 mV (on *c*-cut) and ~107 mV (on *r*-cut) (supplementary material Fig. S1) [27]. Thus, the 40 mV (107 mV-67 mV) difference between the two cases without surface roughness normalization increased to 80 mV (217 mV-137 mV) after the surface roughness normalization, which is the difference of interest for the ongoing discussion.

Remarkably, this observation (i. e. enhanced OER activity even for lower morpho-roughness case) reveals the significance of exposed facets and related surface terminations, and the difference in sheet resistance, suggesting that a change in the surface oxygen binding strength on these high-energetic (111) and (110) SrRuO$_3$ surfaces influences surface oxygen electro-adsorption and corresponding OER activities. Chang *et al.*, also reported that OER activity of SrRuO$_3$ changes with crystallographic orientations as (111) > (110) > (001) [4]. Thus, the combination of lower electrical resistance and higher density of active (Ru) sites and concurrent high surface energy of (111) SrRuO$_3$ film on *c*-Al$_2$O$_3$ in contrast to (110) SrRuO$_3$ film on *r*-Al$_2$O$_3$ sapphire are responsible for better catalytic performance.

To explore the possible role of active sites, we then grew four 20 nm pillars of the SrRuO$_3$ film on the top of uniform 20 nm SrRuO$_3$ film on the $c$-Al$_2$O$_3$ and $r$-Al$_2$O$_3$ films by mechanical masking, as shown in the schematics (Fig. 3(a), (b)). The step-textured nature have a uniform size with average spacing of ~1 mm (supplementary material Fig. S2) [27]. The OER data of step-textured films were compared with 40 nm uniform films grown on two different crystal facets (Fig. 3(c)-(f)). As seen, the reaction started taking place at much lower potential for $c$-plane than $r$-plane, clearly establishing the role surface termination. Interestingly, we observed a remarkable reduction in the overpotential by ~25 mV (on $c$-Al$_2$O$_3$) for the designed geometry (Fig. 3(d). We also ensured that the result is completely reproducible by several tests (two representative cases are compared (supplementary material Fig. S3) [27]. Interestingly, the film grown on $r$-Al$_2$O$_3$ with the same geometry exhibited exactly reverse trend, i.e. higher overpotential (increased by ~15 mV) with respect to the uniform (flat) 40 nm film (Fig. 3(f)). Thus, in addition to the active site effect, the difference in the chemical coordination at the step edges due to change in the local electronic structure is also plays a crucial role in the OER activity.

We also grew several films with different pillar thicknesses (on $c$-Al$_2$O$_3$), and they invariably showed the reduction in the overpotential (Fig. 4(a),(b)), clearly establishing the fact the active sites at the step edges are an important for the enhancement in the OER activity. It is important to note that in mechanically masked film growth, the vertical step is never ideally sharp; hence numerous active step edge sites are invariably present. Moreover, the step edge depth in our case is ~200 Å, which corresponds to ~87 unit cells of (111) and ~71-unit cell of (110). Step edges on such a wall would have high enough step edge density indeed.

In order to probe the surface oxygen binding strength, we also performed the XPS. Interestingly, we observed a moderate shift (~0.5 eV) towards higher energy (reduced electron density) in the XPS for the step-textured film with respect to the uniform film (Fig. 4(c) and supplementary material Fig. S4) [27]. This suggest the reduction in exchange interaction, consequently increasing the local charge on the surface atoms, elucidating the chemical role at the steps of step-textured film [28].

We further analyzed the Tafel plots for both the uniform (flat) film and step-textured film (on $c$-Al$_2$O$_3$) to compare the electrocatalytic activity as well as for elucidating the reaction mechanism of electrocatalysts. The sensitivity of the electric current response to the potential (Tafel slope) provides information associated with the rate determining steps [1,2]. As seen, the corresponding Tafel slope decreases from 81 mV/dec for the uniform (flat) film case to the 75 mV/dec for the step-textured film (Fig. 4(d)). The result suggests faster reaction rate in the case of geometrically engineered step-textured film, i.e. the surface adsorbed species produced in the early stage of the OER remain predominant. This clearly suggests that the step-textured film is kinetically a superior electrocatalyst as compared to the uniform (flat) film.

To further gain insight about the origin of the enhancement in OER on step-textured surface, we performed DFT based electronic structure calculations that helps to elucidate the complete reaction pathway. The electronic structure calculations of correlated metallic compounds are not straightforward; however, the results of our calculations, specially the projected density of states and optical absorption spectra, are in agreement with the optical study of $SrRuO_3$ by Ahn *et al.* [29]. Fig. 5(a) shows the reaction coordinate where the change in free energy ($\Delta G$) is plotted as a function of reaction progress when the reaction intermediates are adsorbed on top of uniform $SrRuO_3$ without step edge geometry. It is observed that the reaction step from OH* → O* formation indicates the energetically favorable half OER pathway while the path step from O* → OOH* formation is found to be energetically unfavorable therefore indicating the reaction hindrance in this step [13, 30]. On the other hand, Fig. 5(b) represents the scenario when a step–textured geometry is created on $SrRuO_3$ surface. In this case, all of the reaction steps i.e. from OH* → O* → OOH* are energetically favorable as far as OER mechanism is concerned. This clearly suggest the enhancement of OER performance for the step-textured films.

## 4. Conclusions

In conclusion, we have shown that geometrically designed (111) $SrRuO_3$ thin film on *c*-$Al_2O_3$ substrate shows enhanced OER activity with the incorporation of active sites, in contrast of an opposite effect for the (110) $SrRuO_3$ film on *r*-$Al_2O_3$ substrates, as a consequence of the presence of active sites with higher energetic polar surface termination. Density functional theory confirms the enhancement with favorable energetic reaction pathway. Our study demonstrates that presence of active sites is indeed a vital parameter for the enhancement in OER activity. This geometrical engineering approach might be a useful step towards using perovskite metal oxide thin films as an active electrocatalyst for the production of renewable energy.

**Declaration of Competing Interest**

The authors declare that they have no competing financial interests that could have appeared to influence the work reported in this paper.

**Acknowledgments**

S. O. would like to acknowledge the funding support by the DST Nanomission Thematic unit (SR/NM/TP-13/2016). We would like to thank Yogeshwar More for the schematics, Dr. Ram Janay Choudhary from UGC-DAE-CSIR Indore for XRD, Dr. Mallikarjuna Rao Motapothula of Uppsala University for the RBS measurements, and Rushikesh Magdum for help in experiments. We would also like to acknowledge NSC for providing the computing time.

**References**

[1] N. –T. Suen, S. –F. Hung, Q. Quan, N. Zhang, Y.-J. Xu, H. M. Chen, **Electrocatalysis for the oxygen evolution reaction: recent development and future perspectives**, Chem. Soc. Rev. 46 (2017) 337. (https://doi.org/10.1039/C6CS00328A)

[2] C. Hu, L. Zhang, J. Gong, **Recent progress made in the mechanism comprehension and design of electrocatalysts for alkaline water splitting**, Energy Enviorn. Sci. 12 (2019) 2620. (https://doi.org/10.1039/C9EE01202H)

[3] F. Song, L. Bai, A. Moyasiadou, S. Lee, C. Hu, L. Lairdet, X. Hu, **Transition Metal Oxides as Electrocatalysts for the Oxygen Evolution Reaction in Alkaline Solutions: An Application-Inspired Renaissance**, J. Am. Chem. Soc. 140 (2018) 7748. (https://doi.org/10.1021/jacs.8b04546)

[4] S. H. Chang, N. Danilovic, K.-C. Chang, R. Subbraman, A. P. Paulikas, D. D. Fong, M. J. Highland, P. M. Baldo, V. R. Stamenkovic, J. W. Freeland, J. A. Eastman, M. Markovic, **Functional links between stability and reactivity of strontium ruthenate single crystals during oxygen evolution**, Nat. Commun. 4 (2014) 4191. (https://doi.org/10.1038/ncomms5191)

[5] S. Hirai, T. Ohno, R. Uemura, T. Maruyama, M. Furunaka, R. Fukunaga, W. –T. Chen, H. Suzuki, T. Matsuda, S. Yagi, **$Ca_{1-x}Sr_xRuO_3$ perovskite at the metal–insulator**

[6] L. C. Seitz, C. F. Dickens, K. Nishio, Y. Hikita, J. Montoya, A. Doyle, C. Kirk, A. Vojvodic. H. Y. Hwang, J. K. Norskov, T. F. Jaramillo, **A highly active and stable IrO$_x$/SrIrO$_3$ catalyst for the oxygen evolution reaction**, Science 353 (2016) 6303. (DOI: 10.1126/science.aaf5050)

[7] J. R. Petrie, V. R. Cooper, J. W. Freeland, T. L. Meyer, Z. Zhang, D. A. Lutterman, H. N. Lee, **Enhanced Bifunctional Oxygen Catalysis in Strained LaNiO$_3$ Perovskites**, J. Am. Ceram. Soc. 138 (2016) 2488. (https://doi.org/10.1021/jacs.5b11713)

[8] J. Schloz, M. Risch, G. Wartner, C. Luderer, V. Roddatis, C. Jooss, **Tailoring the Oxygen Evolution Activity and Stability Using Defect Chemistry**, Catalysis 7 (2017) 139. (https://doi.org/10.3390/catal7050139)

[9] K. A. Stoerzinger, W. S. Choi, H. Jeen, H. N. Lee, Y. Shao-Horn, **Role of Strain and Conductivity in Oxygen Electrocatalysis on LaCoO$_3$ Thin Films**, J. Phys. Chem. Lett. 6 (2015) 487. (https://doi.org/10.1021/jz502692a)

[10] Y. Matsumoto, H. Manabe, E. Sato, **Oxygen Evolution on La$_{1-x}$Sr$_x$CoO$_3$ Alkaline Solutions**, J. Electrochem. Soc. 4 (1980) 811. (10.20964/2016.10.01)

[11] J. Suntivich, K. J. May, H. Gasteiger, J. B. Goodeough, Y. Shoa-Horn, **A Perovskite Oxide Optimized for Oxygen Evolution Catalysis from Molecular Orbital Principles**, Science 334 (2011) 1383. (DOI: 10.1126/science.1212858)

[12] J. O'M. Beckris T. Otagawa, **Mechanism of Oxygen Evolution on Perovskites**, J. Phys. Chem. 87 (1983) 2960. (https://doi.org/10.1021/j100238a048)

[13] I. C. Man, H-Y. Su, F. Calle-Vallejo, H. A. Hansen, J. I. Martinez, N. G. Inoglu, J. Kitchin, T. F. Jaramillo, J. K. Nørskov, J. Rossmeisl, **Universality in Oxygen Evolution Electrocatalysis on Oxide Surfaces**, ChemCatChem. 3 (2011) 1159. (https://doi.org/10.1002/cctc.201000397)

[14] T. Reier, M. Oezaslan, P. Strasser, **Electrocatalytic Oxygen Evolution Reaction (OER) on Ru, Ir, and Pt Catalysts: A Comparative Study of Nanoparticles and Bulk Materials**, ACS Catal. 2 (2012) 1765. (https://doi.org/10.1021/cs3003098)

[15] G. Koster, L. Klein, W. Siemons, G. Rijnders, J. S. Dodge, C.-B. Eom, D. H. A. Blank, M. R. Beasley, **Structure, physical properties, and applications of SrRuO$_3$ thin films**, Rev. Mod. Phys. 84 (2012) 253. (https://doi.org/10.1103/RevModPhys.84.253)

[16] B. J. Kim, D. F. Abbott, X. Cheng, E. Fabbri, M. Nachtegaal, F. Bozza, I. E. Castelli, D. Lebedev, R. Schäublin, C. Copéret, T. Graule, N. Marzari, T. J. Schmidt,

**Unraveling Thermodynamics, Stability, and Oxygen Evolution Activity of Strontium Ruthenium Perovskite Oxide**, ACS Catal. 7 (2017) 3245. (https://doi.org/10.1021/acscatal.6b03171)

[17] A. R. Akbashev, L. Zhang, J. T. Mefford, J. Park, B. Butz, H. Liftman, W. C. Chueh, A. Vojvodic, **Activation of ultrathin SrTiO$_3$ with subsurface SrRuO$_3$ for the oxygen evolution reaction**, Energy Environ. Sci. 11 (2018) 1762. (https://doi.org/10.1039/C8EE00210J)

[18] S. A. Lee, S. Oh, J.-Y. Hwang, M. Choi, C. Youn, J. W. Kim, S. H. Chang, S. Woo, J.-S. Bae, S. Park, Y.-M. Kim, T. Choi, S. W. Kim, W. S. Choi, **Enhanced electrocatalytic activity via phase transitions in strongly correlated SrRuO$_3$ thin films**, Energy Environ. Sci. 10 (2017) 924. (https://doi.org/10.1039/C7EE00628D)

[19] M. Retuerto, L. Pascual, F. Calle-Vallejo, P. Ferrer, D. Gianolio, A. M. Pereira, Á. García, J. Torrero, M. T. Fernández-Díaz, P. Bencok, M. A. Peña, J. L G. Fierro, S. Rojas, **Na-doped ruthenium perovskite electrocatalysts with improved oxygen evolution activity and durability in acidic media**, Nat. Commun. 10 (2019) 2041. (https://doi.org/10.1038/s41467-019-09791-w)

[20] W. Kohn and L. J. Sham, **Self-Consistent Equations Including Exchange and Correlation Effects**, Phys. Rev. 140 (1965) A1133. (https://doi.org/10.1103/PhysRev.140.A1133)

[21] G. Kresse, J. Furthmüller, **Efficient iterative schemes for ab initio total-energy calculations using a plane-wave basis set**, Phys. Rev B 54 (1996) 11169. (https://doi.org/10.1103/PhysRevB.54.11169)

[22] J. P. Perdew, K. Burke, M. Ernzerhof, **Generalized Gradient Approximation Made Simple**, Phys. Rev. Lett. 77 (1996) 3865. (https://doi.org/10.1103/PhysRevLett.77.3865)

[23] P. E. Blochl, **Projector augmented-wave method**, Phys. Rev. B 50 (1994) 17953. (https://doi.org/10.1103/PhysRevB.50.17953)

[24] H. J. Monkhorst, J. D. Pack, **Special points for Brillonin-zone integrations**, Phys. Rev. B 13 (1976) 5188. (https://doi.org/10.1103/PhysRevB.13.5188)

[25] S. Grimme, Semiempirical **GGA-Type Density Functional Constructed with a Long-Range Dispersion Correction**, J. Comput. Chem. 27 (2006) 1787. (https://doi.org/10.1002/jcc.20495)

**[26]** A. J. Grutter, F. J. Wong, E. Arenholz, A. Vailionis, Y. Suzuki, **Evidence of high-spin Ru and universal magnetic anisotropy in SrRuO$_3$ thin films**, Phys. Rev. B 85 (2012) 134429. (https://doi.org/10.1103/PhysRevB.85.134429)

**[27]** See the supplementary material for more information about the morphological, chemical, and electrocatalytic characterizations.

**[28]** D. –Y. Kuo, C. J. Eom, J. K. Kawasaki, G. Petretto, J. N. Nelson, G. Hautier, E. J. Crumlin, K. M. Shen, D. G. Schlom, J. Suntivich, **Influence of Strain on the Surface−Oxygen Interaction and the Oxygen Evolution Reaction of SrIrO$_3$**, J. Phys. Chem. C, 122 (2018) 4359. (https://doi.org/10.1021/acs.jpcc.7b12081)

**[29]** J. S. Ahn, J. Bak, H. S. Choi, T. W. Noh, J. E. Han, Y. Bang, J. H. Cho, and Q. X. Jia, **Spectral Evolution in (Ca,Sr)RuO$_3$ near the Mott-Hubbard Transition**, Phys. Rev. Lett. 82 (1999) 5321. (https://doi.org/10.1103/PhysRevLett.82.5321)

**[30]** B. Garlyyev, J. Fichtner, O. Pique, O. Schneider, A. S. Bandarenka, F. Calle-Vallejo, **Revealing the nature of active sites in electrocatalysis**, Chem. Sci. 10 (2019) 8060. (https://doi.org/10.1039/C9SC02654A)

**FIGURE CAPTIONS**

**Fig. 1. (Color online)** (a)-(b) X-ray diffraction, (c) surface morphology and (d) sheet resistance of SrRuO$_3$ thin films grown on $c$-and $r$-Al$_2$O$_3$ substrate. XRD shows growth along (111) and (110) for films on $c$-Al$_2$O$_3$ and $r$-Al$_2$O$_3$ substrate, respectively.

**Fig. 2. (Color online)** (a)-(b) Schematic representation of crystal structures (left panel) and two possible ideal polar termination layers (right panel) from a crystallographic point of view for (111) and (110) oriented SrRuO$_3$. (c) Electrocatalytic OER performance and (d) comparison of overpotential of uniform (flat, non-textured) SrRuO$_3$ films grown on $c$-Al$_2$O$_3$ and $r$-Al$_2$O$_3$. The current density was normalized duly with the respective film's surface roughness value.

**Fig. 3. (Color online)** Schematics displaying (a) uniform (flat, non-textured), and (b) step-textured SrRuO$_3$ thin films (dark yellow) grown on sapphire (Al$_2$O$_3$) substrates (grey) used for the OER measurements. The electrocatalytic OER performance, overpotential (measured at $j$ = 1 mA/cm$^2$) values for uniform 40 nm (flat, non-textured) and step-textured (20 nm uniform surface + 20 nm pillars) films on (c-d) $c$-Al$_2$O$_3$, and (e-f) $r$-Al$_2$O$_3$ substrates.

**Fig. 4. (Color online)** (a) Comparison of electrocatalytic OER performance, and (b) overpotential (measured at $j$ = 1 mA/cm$^2$) for different bottom layer and step thicknesses for the step-textured film (on $c$-Al$_2$O$_3$) with respect to the uniform 40 nm film (flat, non-textured). (c) XPS spectra at the Ru-3$d$ edge and (d) Tafel plots for uniform (flat, non-textured) 40 nm film and step-textured film (20 nm uniform surface + 20 nm pillars) on $c$-Al$_2$O$_3$.

**Fig. 5. (Color online)** Calculated free energies of the OER intermediates for the (a) uniform and (b) step-textured SrRuO$_3$ film (atom colors: Sr-blue, Ru-red and O-green).

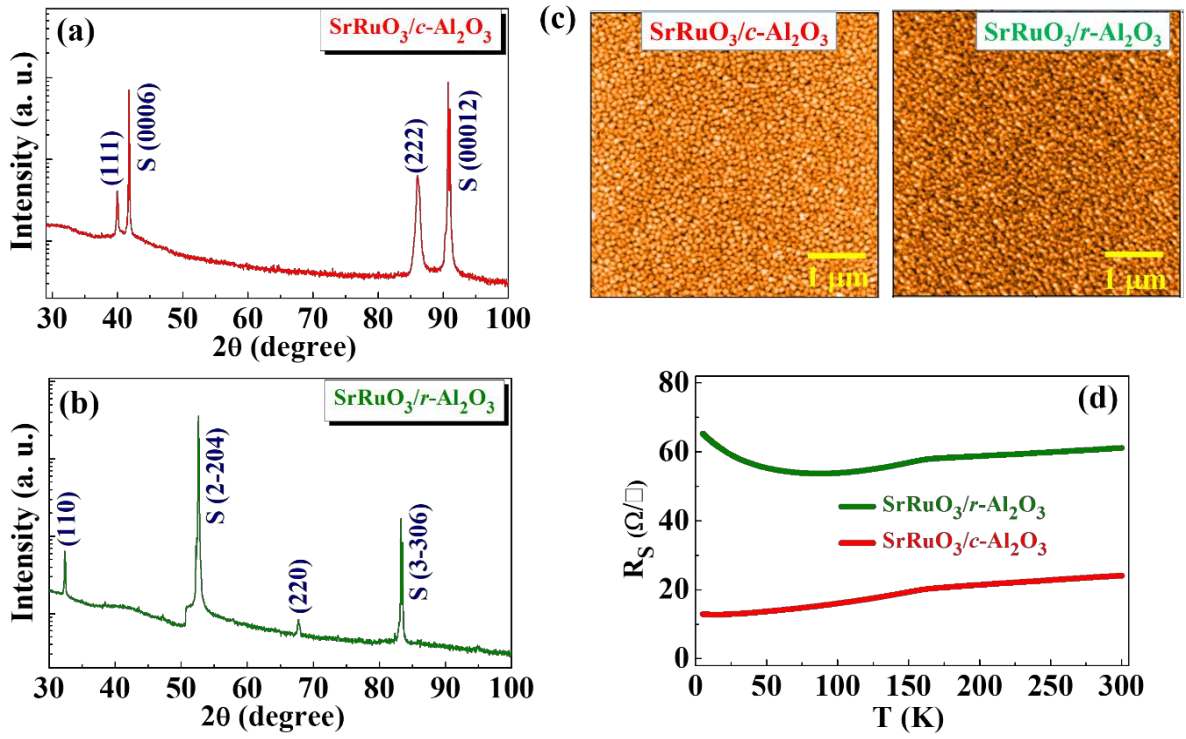

**(FIGURE-1)**

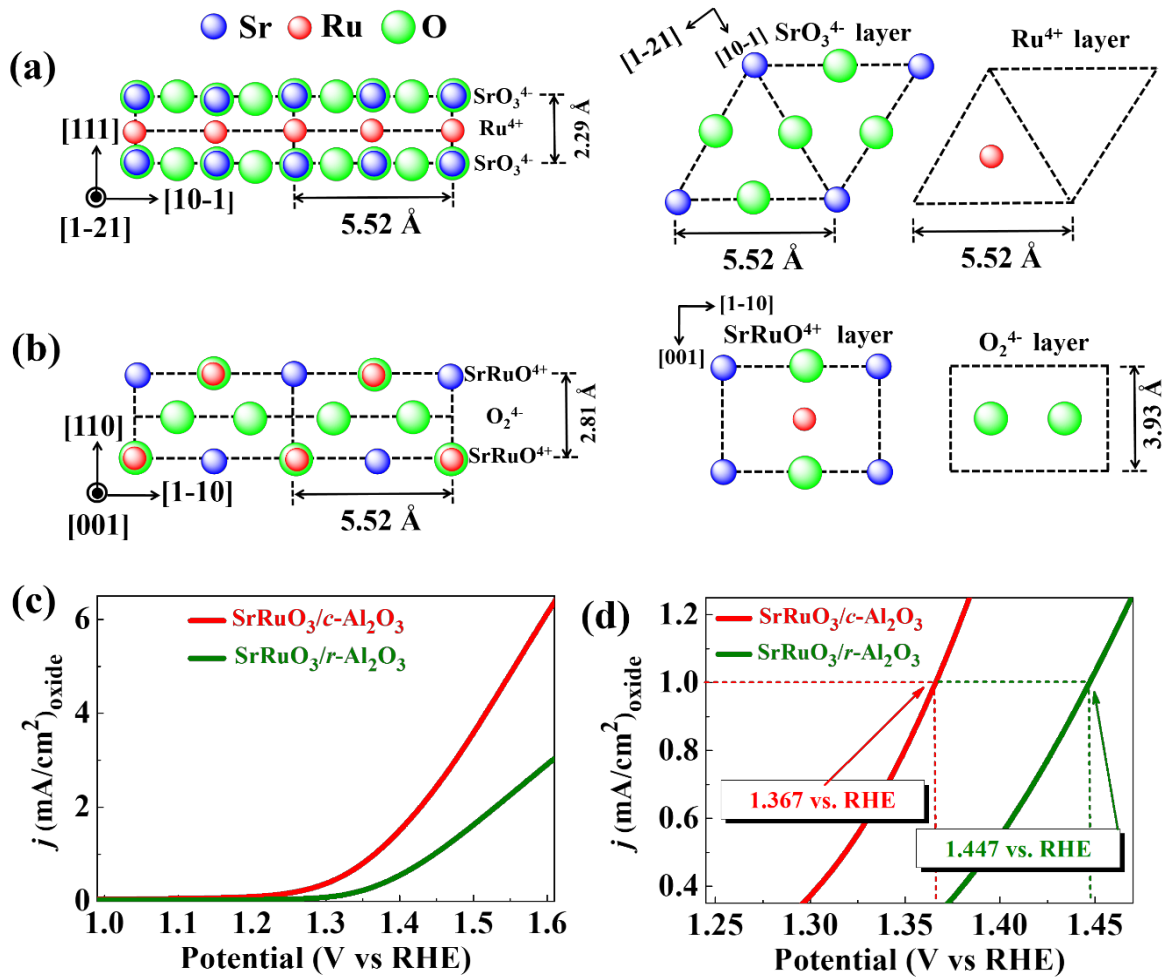

**(FIGURE-2)**

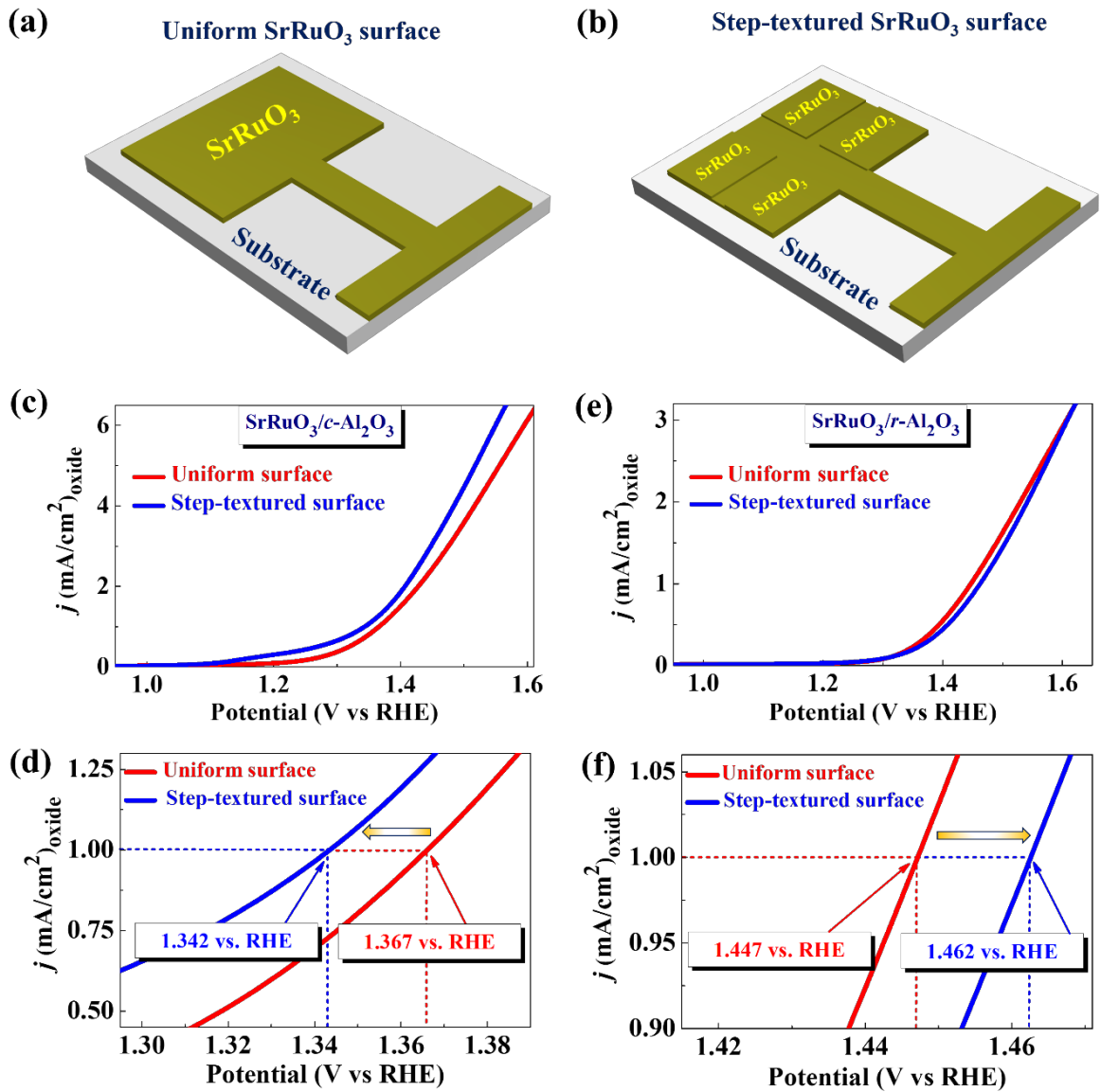

**(FIGURE-3)**

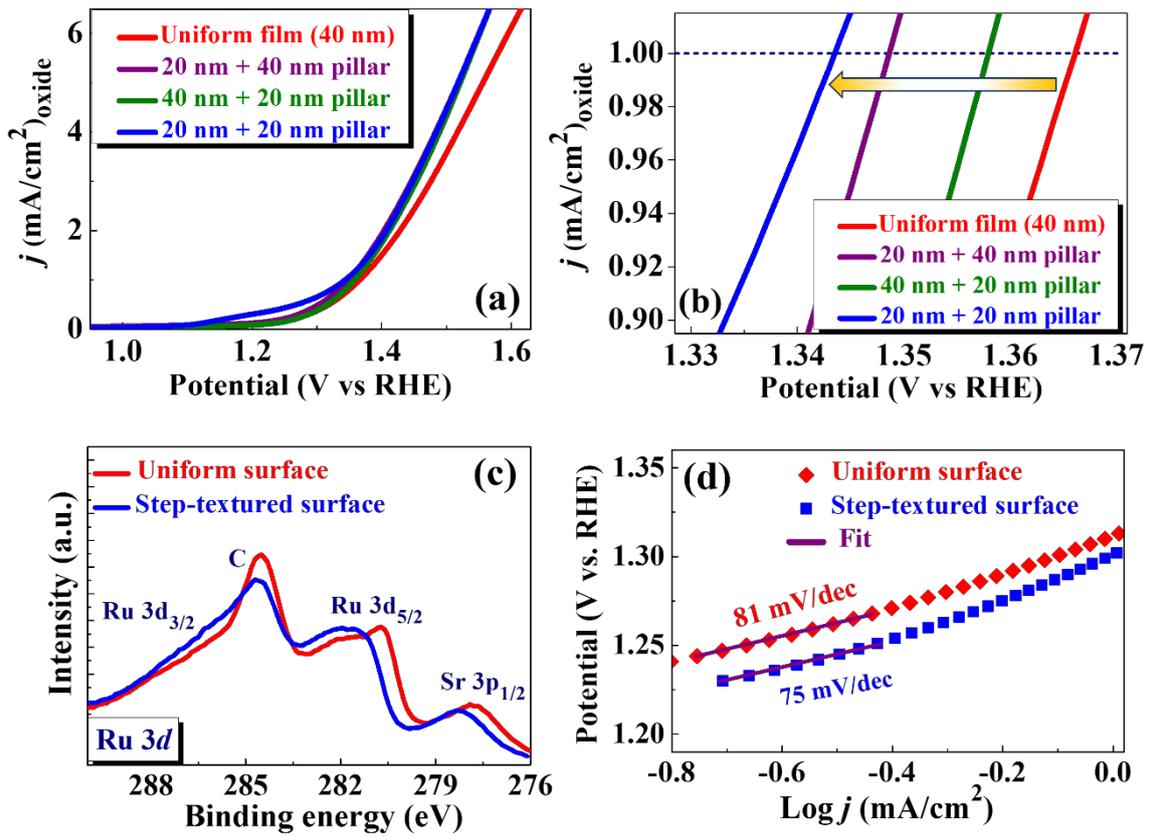

**(FIGURE-4)**

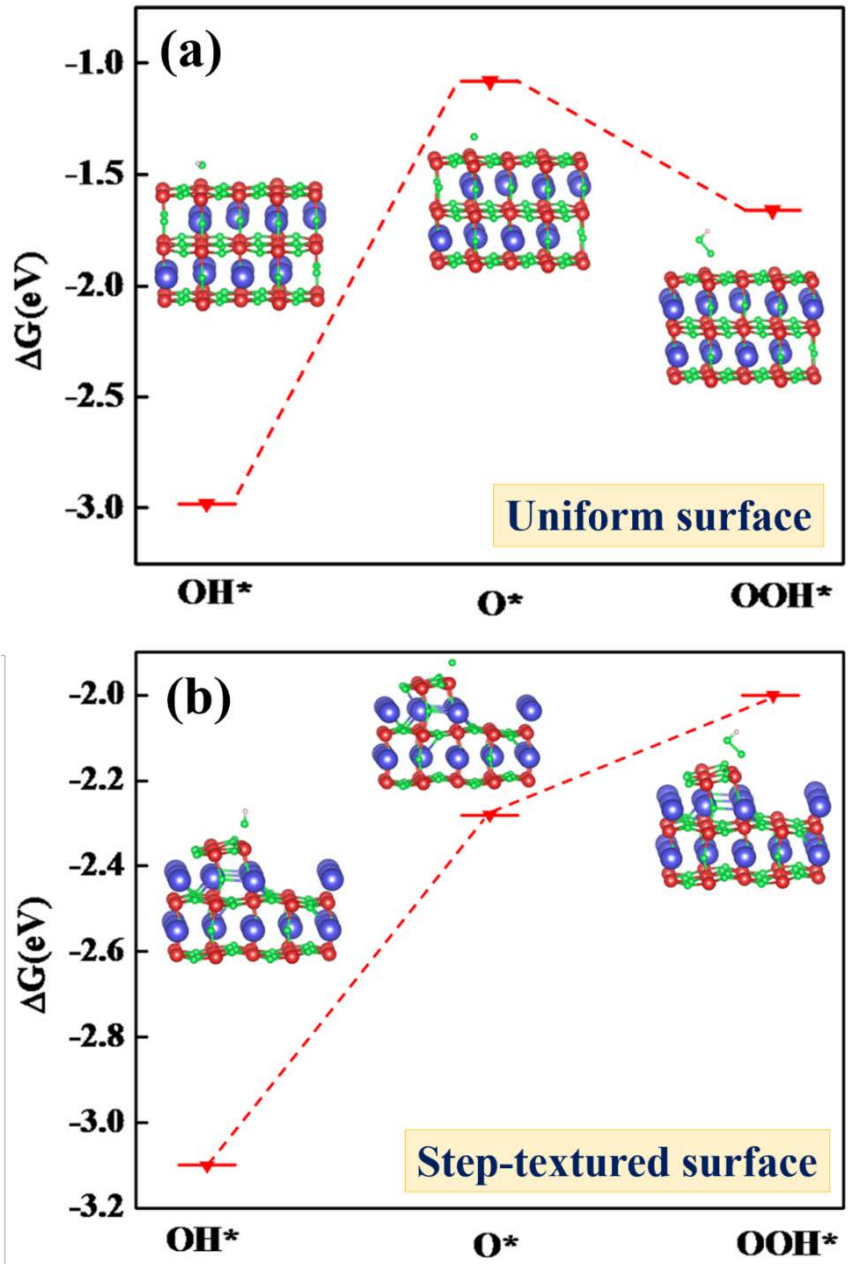

**(FIGURE-5)**